\documentclass[prb,aps,twocolumn,superscriptaddress,lengthcheck,showpacs,amsmath,amssymb,floatfix,longbibliography]{revtex4-1}
\usepackage{graphicx}  
\usepackage{hyperref}
\usepackage{microtype}
\usepackage{float}

\hypersetup{colorlinks,linkcolor=blue,citecolor=blue,urlcolor=blue}
\def\equationautorefname~#1\null{Eq.~(#1)\null}

\begin{document}

\title{A chiral route to spontaneous entanglement generation}
\author{Carlos Gonzalez-Ballestero}
\email{carlos.ballestero@uam.es}
\affiliation{Departamento de F{\'\i}sica Te{\'o}rica de la Materia Condensada and Condensed Matter Physics Center (IFIMAC), Universidad Aut\'onoma de Madrid, E-28049 Madrid, Spain}

\author{Alejandro Gonzalez-Tudela}
\affiliation{Max-Planck-Institut f\"{u}r Quantenoptik Hans-Kopfermann-Str. 1. 85748 Garching, Germany}

\author{Francisco J. Garcia-Vidal}
\affiliation{Departamento de F{\'\i}sica Te{\'o}rica de la Materia Condensada and Condensed Matter Physics Center (IFIMAC), Universidad Aut\'onoma de Madrid, E-28049 Madrid, Spain}
\affiliation{Donostia International Physics Center (DIPC), E-20018 Donostia/San Sebastian, Spain}

\author{Esteban Moreno}
\affiliation{Departamento de F{\'\i}sica Te{\'o}rica de la Materia Condensada and Condensed Matter Physics Center (IFIMAC), Universidad Aut\'onoma de Madrid, E-28049 Madrid, Spain}


\begin{abstract}
We study the generation of spontaneous entanglement between two qubits chirally coupled to a waveguide. The maximum achievable concurrence is demonstrated to increase by a factor of $4/e \sim 1.5$ as compared to the non-chiral coupling situation. The proposed entanglement scheme is shown to be robust against variation of the qubit properties such as detuning and separation, which are critical in the non-chiral case. This result relaxes the restrictive requirements of the non-chiral situation, paving the way towards a realistic implementation. Our results demonstrate the potential of chiral waveguides for quantum entanglement protocols.
\end{abstract}

\maketitle


Efficient quantum circuits are a very important ingredient for the development of quantum computing \cite{Kimble}. Usual implementations of these devices require platforms where information, usually in the form of photons, can be easily introduced and extracted \cite{OBrien07122007}. Several systems based on waveguides have been proposed for quantum circuitry, from superconducting stripes  \cite{NiemczykNATPHYS} to dielectric \cite{SLOTPhysRevA,PolzikPRL2014}, photonic crystal \cite{Wong200947,KIMBLEnatcomm2014} or plasmonic waveguides \cite{LukinNATURE}. In this context, the interaction between the guided photons and quantum emitters is critical for various processes such as the creation of entangled states between the qubits. Spontaneous entanglement generation in waveguide setups has already been predicted \cite{Fleischhauer,PRLDiego}. Many other interesting phenomena like mesoscopic entanglement \cite{TudelaPORRAS}, long-distance quantum beats \cite{BarangerPRL}, or the formation of sub- and superradiant states \cite{WallraffSCIENCE}, show that waveguides are excellent platforms for quantum information processing.

Recently, systems of emitters chirally coupled to waveguides have attracted a lot of attention both theoretically and experimentally \cite{PRLZoller,PRAzoller,RauschenPRA}. In these configurations, an adequate engineering of the waveguide can be used to break the emission symmetry of the qubits, channelling the emitted photons preferentially into one of the two directions of the waveguide. Chirality in emitter-waveguide coupling is a general effect associated to the so-called spin-orbit interaction of light \cite{RauschenbeutelSCIENCE}. Besides theoretical studies, many experiments have reported chiral emission with a very large degree of directionality, from nanoparticles and atomic ensembles in dielectric waveguides \cite{RauschenbeutelSCIENCE,RauschenbeutelNATCOMM} to quantum dots in nanobeams \cite{COLEsarxiV} and photonic crystals \cite{OULTONsarxiV,KUIPERSsarxiV,LodahlarxiV3}. Among these setups, the latter turn out to be especially promising systems as they combine large directionalities of around $90\%$ with high emitter-waveguide coupling fractions (up to $98\%$). Consequently, they have been proposed as ideal platforms for implementation of quantum logical gates \cite{LodahlarxiV3}.

\begin{figure}
	\centering
	\includegraphics[scale=0.2]{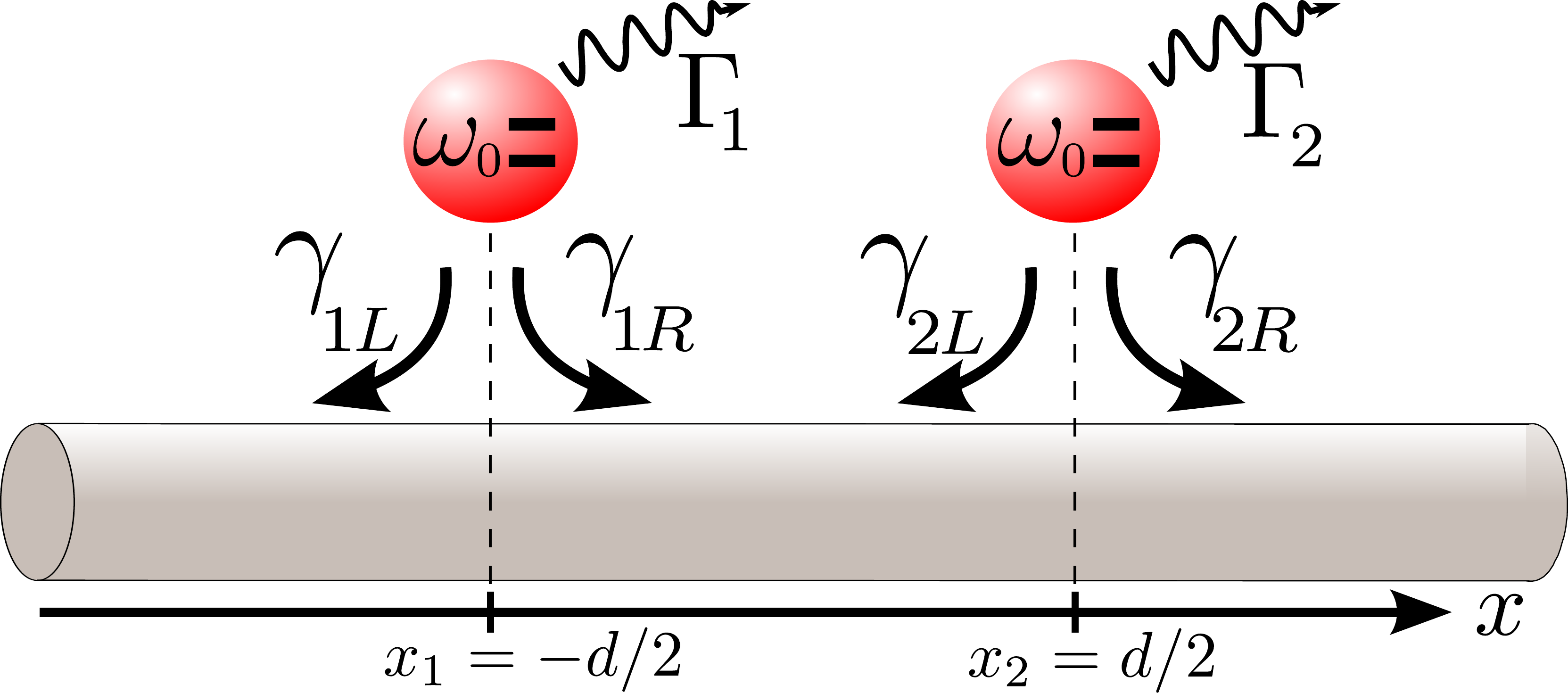}
	\caption{System under study. Two qubits of frequency $\omega_0$ and separated by a distance $d$ are placed in the vicinities of a waveguide. The energies $\gamma_{j\alpha}$ quantify the chiral coupling of qubit $j$ to the photonic propagating mode $\alpha ~ (=L,R)$. In the same fashion, the decay rate into the 3D environment and other lossy modes is described by a coupling constant $\Gamma_j$.}\label{figg1}
	\vspace{-0.3cm}
\end{figure}

In this Rapid Communication, we tackle the problem of spontaneous entanglement generation between two qubits chirally coupled to a waveguide. In the first part of this work, we present an analytical solution to the Master Equation describing the evolution of the system state, showing how chirality allows for an enhancement up to $\sim 50\%$ in the maximum generated entanglement as compared to the non-chiral case. In the second part of the paper, we present a more complete formalism in which non-Markovian effects are explicitly accounted for by fully diagonalizing the Hamiltonian in the single-excitation subspace. We use this formalism to 
demonstrate the robustness of the entanglement generation scheme against the detuning between the qubits, the total coupling rate, and the qubit-qubit separation.

The system under study is depicted in \autoref{figg1}. Two emitters $1$ and $2$, modelled as two-level systems of frequency $\omega_0$, are coupled to the propagating photonic modes of a waveguide. The emitters are separated by a distance $d = x_2 - x_1$, and coupled to the right and left propagating photons through the energy constants $\gamma_{jR}$ and $\gamma_{jL}$ ($j=1,2$), respectively. In a chiral coupling scheme such as the one analyzed in this work, these constants are different $(\gamma_{jR} \neq \gamma_{jL})$. Other deexcitation processes into free space or additional lossy modes are taken into account through the decay rates $\Gamma_j$. The three coupling constants of each qubit are used to define a usual figure of merit in waveguide systems, namely the coupling fraction or beta factor, given by $\beta_j = (\gamma_{jR} + \gamma_{jL})/(\gamma_{jR} + \gamma_{jL} + \Gamma_j)$.

Our aim is to analyze spontaneous entanglement generation when qubit $1$ (the qubit on the left) is initially excited. We start by solving the system dynamics, obtaining the time evolution of the reduced density matrix of the two-qubit subsystem, $\rho$. In the first part of this work, we will follow a usual approach undertaken in quantum optics, in which the problem is simplified by tracing out the photonic degrees of freedom under the so-called Markov approximation \cite{Louisell}. In this situation, the only dynamical variable is the density matrix $\rho$, whose evolution is governed by the following general Master Equation:
\begin{equation} \label{MEQ}
\begin{split}
\dot{\rho} = -i\left[H,\rho\right]+ \sum_{j=1,2} \gamma_j \mathcal{L}_{\sigma_j,\sigma_j} [\rho]+&\\
+ \sqrt{\gamma_{2R}\gamma_{1R}} \mathcal{L}_{\sigma_2,\sigma_1}[\rho] +& \sqrt{\gamma_{2L}\gamma_{1L}}\mathcal{L}_{\sigma_1,\sigma_2}[\rho] ,
\end{split}
\end{equation}
where we define $\gamma_j \equiv \left(\gamma_{jR} + \gamma_{jL}\right)/2$. The generalized Lindblad superoperators, $\mathcal{L}_{\sigma_a,\sigma_b}$, are employed to describe incoherent processes, in this case a waveguide-mediated interaction. They are expressed as
\begin{equation}
		\mathcal{L}_{\sigma_a,\sigma_b}[\rho] =\left(e^{-i2\pi D_{ab}}[\sigma_a,\rho \sigma_b^\dagger]-e^{i2\pi D_{ab}}[\sigma_a^\dagger,\sigma_b\rho]\right)\!,
\end{equation}
 where $D_{ab}\equiv \vert x_a - x_b\vert/\lambda_0$, $\lambda_0 = 2\pi v_g/\omega_0$ is the emission wavelength of the qubits, and $v_g$ is the group velocity of the guided photons. The bare Hamiltonian of the system is given by $H = \omega_0(\sigma_1^\dagger \sigma_1+\sigma_2^\dagger\sigma_2 )$, where $\sigma_j$ is the annihilation operator of qubit $j$. Note that \autoref{MEQ} is particular of chiral configurations \cite{PRAzoller}, and is reduced to its more common form in the case $\gamma_{jR} = \gamma_{jL}$. In order to focus on the fundamental aspects of the chiral system, we will first particularize our study to the lossless case (i.e., $\beta_j = 1$), including the losses in the second part of the work.

Combining \autoref{MEQ} with the particular initial conditions of our problem, $\rho(0) = \sigma_1^\dagger \vert 0\rangle\langle 0 \vert \sigma_1$, and expressing $\rho$ in the usual basis $\lbrace \vert 0 \rangle,\vert 1\rangle \equiv\sigma_1^\dagger \vert 0\rangle, \vert 2\rangle \equiv\sigma_2^\dagger \vert 0\rangle,\vert 3\rangle \equiv\sigma_1^\dagger\sigma_2^\dagger \vert 0\rangle\rbrace$, the only non-zero elements of the density matrix are the populations $\rho_{00},\rho_{11},\rho_{22},$ and the coherence $\rho_{12}$. Three of these quantities are coupled through the following system of differential equations,
\begin{equation}
\begin{split}
\dot{\rho}_{11}& =\\ &-2\gamma_1\rho_{11} -  \sqrt{\gamma_{1L}\gamma_{2L}}\left(e^{i2\pi \tilde{d}}\rho_{21}+e^{-i2\pi \tilde{d}}\rho_{12}\right),
\end{split}
\end{equation}
\vspace{-0.5cm}
\begin{equation}
\begin{split}
\dot{\rho}_{22} &=\\& -2\gamma_2\rho_{22} -  \sqrt{\gamma_{1R}\gamma_{2R}}\left(e^{i2\pi \tilde{d}}\rho_{12}+e^{-i2\pi \tilde{d}}\rho_{21}\right),
\end{split}
\end{equation}
\vspace{-0.4cm}
\begin{equation}
\begin{split}
\dot{\rho}_{12} = -(&\gamma_1 + \gamma_2)\rho_{12} \\ -&\sqrt{\gamma_{1R}\gamma_{2R}}\rho_{11}e^{-i2\pi \tilde{d}}- \sqrt{\gamma_{1L}\gamma_{2L}}\rho_{22}e^{i2\pi \tilde{d}},
\end{split}
\end{equation}
with the normalized distance $\tilde{d} = d/\lambda_0$. Once these equations are solved, we can compute the qubit-qubit entanglement which we quantify with the Wootters concurrence $C$ \cite{Wootters}, a widely used measure thanks to its simple calculation and its intuitive bounds. Indeed, the concurrence ranges from $0$ for non entangled states, to $1$ for maximally entangled configurations. A straightforward calculation demonstrates that in this case, $C = 2\vert \rho_{12}\vert$. An analytical solution of \autoref{MEQ} can be obtained when the two qubits are equally coupled, i.e., $\gamma_1 = \gamma_2 =\gamma$. The following expression for the concurrence is obtained:
\begin{equation}\label{conc}
\begin{split}
C^2(t)& =\sqrt{\frac{(1+\Delta_1)(1+\Delta_2)}{(1-\Delta_1)(1-\Delta_2)}} e^{-4\gamma t} \times \\& \!\!\!\!\!\!\!\!\times\!\Big(\!\sin^2\!\left[2q\gamma t\sin(2\pi \tilde{d})\right]\!+\sinh^2\!\left[2q\gamma t\cos(2\pi \tilde{d})\right]\!\Big),
\end{split}
\end{equation} 
where $q\equiv\left(1-\Delta_1^2\right)^{1/4}\left(1-\Delta_2^2\right)^{1/4}$, and we have introduced the \textit{directionality} of qubit $j$ as the adimensional ratio $\Delta_j = \left(\gamma_{jR}-\gamma_{jL}\right)/\left(\gamma_{jR}+\gamma_{jL}\right)$. 

\begin{figure}
	\centering
	\includegraphics[width=\linewidth]{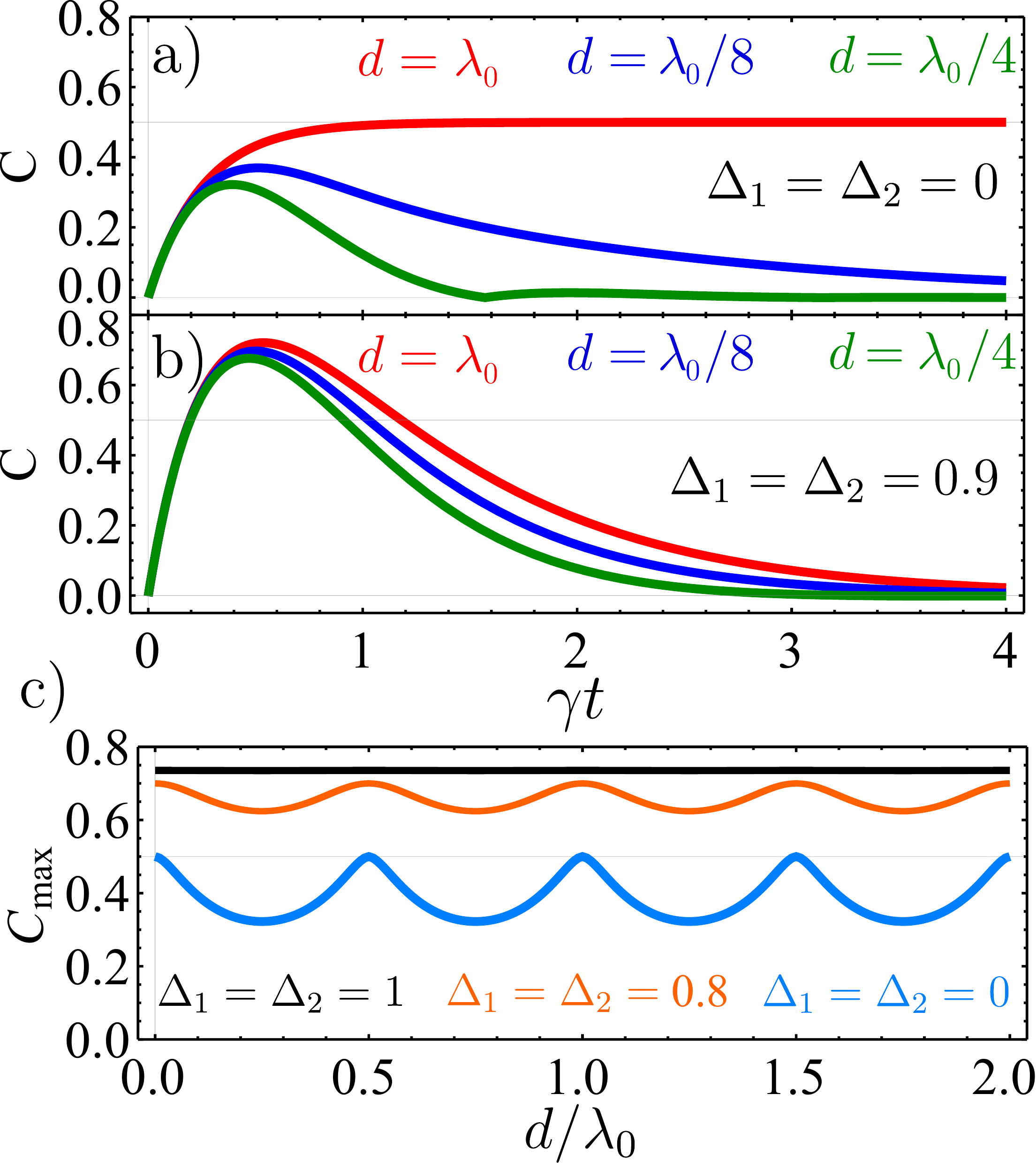}
	\caption{\!Time evolution of the concurrence, \autoref{conc}, for different qubit-qubit separations $d$, in the non-chiral (a) and chiral (b) cases.  c) Dependence of $C_{\text{max}}$ on the separation between the qubits, $d$. The condition $\Delta_1 = \Delta_2$ is chosen because it optimizes the maximum achievable concurrence $C_{\text{max}}$ (see main text). In all three panels the ideal case $\beta_j \!= \!\!1$ is considered.}\label{figg2}
\end{figure}

In order to have a clear view of the advantages brought about by chirality, let us first analyze the non-chiral case, $\Delta_j = 0$. The time evolution of the concurrence in this situation is displayed in \autoref{figg2}a. Regardless of the separation between the qubits, the concurrence is $0$ for $t=0$ since the initial state is separable, and grows up to a maximum value as, due to the exchange of population through the waveguide modes, the state of the qubits becomes entangled. For $t\to \infty$, the behavior of the concurrence depends on the separation $d$. For almost all separations, the population abandons the vicinities of the emitters in the form of propagating photons and, as a consequence, the concurrence decays to zero. However, when the qubits are identical and for very specific separations, $2\tilde{d} = 0,1,2,...$, the behavior is different, as the concurrence not only reaches its maximum achievable value, $C = 0.5$, but also has an infinite lifetime. This peculiar time evolution is caused by the appearance of a Fabry-Perot-like resonance between the qubits, where a photon of energy $\omega_0$ can be trapped forming a standing wave \cite{NJP}. The presence of this localized photon is linked to the fact that the transmittance of one qubit is strictly zero for an incoming resonant photon \cite{FANopticsLetters}. Note that, in a realistic case $(\beta < 1)$ the concurrence always decays with time \cite{PRLDiego}.

When the coupling is chiral, on the other hand, a straightforward calculation demonstrates that the above-mentioned standing wave does not appear \cite{suppl}, as the chirality effectively couples the right- and left- propagating modes and, consequently, the single qubit transmittance never vanishes. As figure \autoref{figg2}b shows, this results in a different time evolution in which two important features arise, namely the weak dependence on the separation $d$, and the achievement of much larger values for the concurrence as compared to the non-chiral situation. The maximum concurrence achieved during the time evolution, $C_{\text{max}}$, is displayed in \autoref{figg2}c as a function of $d$. Clearly, the maximally chiral configuration $\Delta_j = 1$ optimizes the entanglement generation scheme, as $C_{\text{max}}$ reaches a maximum value which, additionally, is independent on $d$. Chirality thus represents an advantage towards a realistic implementation, as it can overcome the critical dependence on the separation in non-chiral configurations. 

\begin{figure}
	\centering
	\includegraphics[scale=0.2]{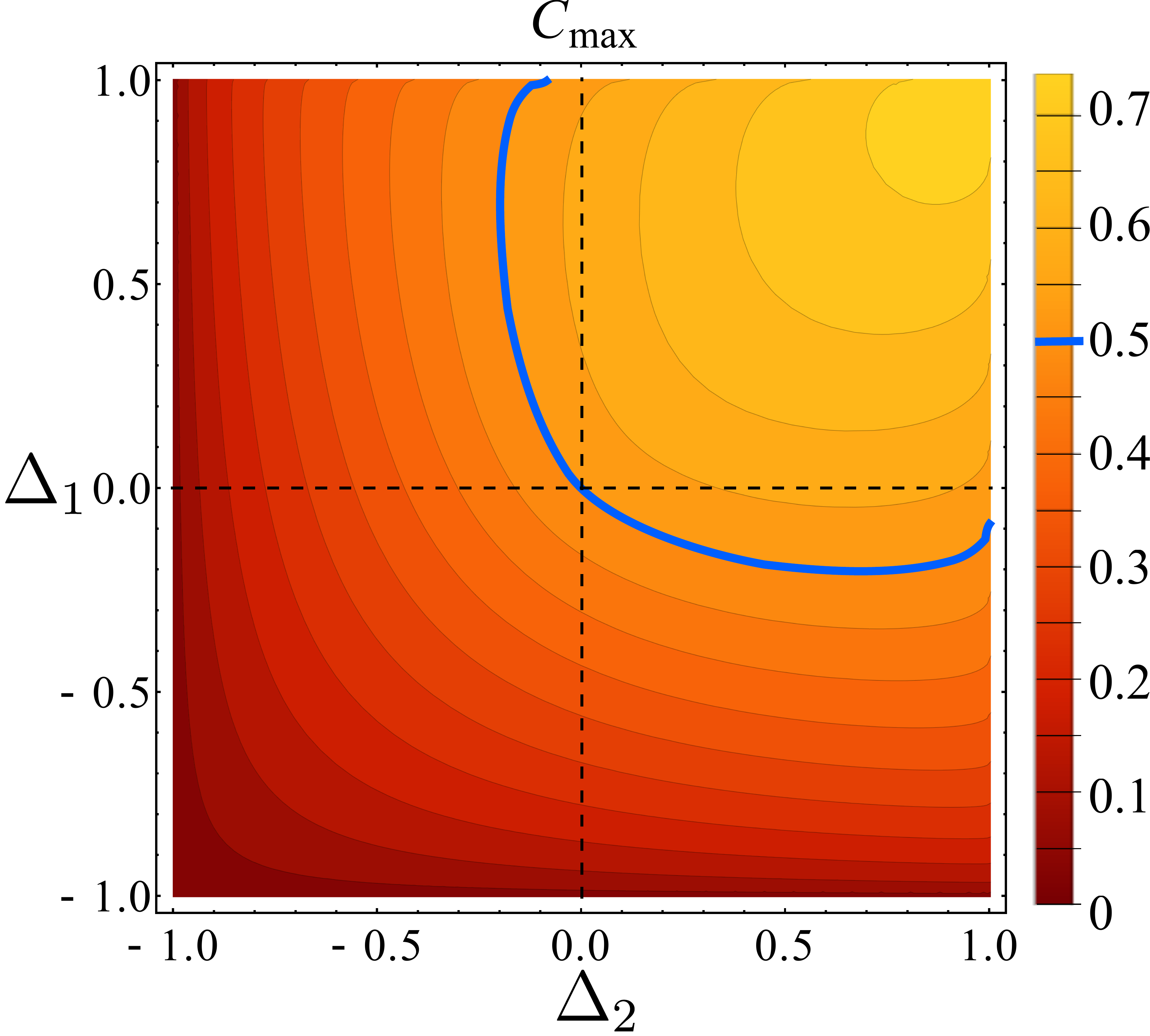}
	\caption{Maximum achievable concurrence in the ideal case $\left(\beta = 1\right)$, versus the directionalities of each qubit, $\Delta_1, \Delta_2$, for the left qubit initially excited, and a separation $d = \lambda_0$. The blue line displays the non-chiral value $C_{\text{max}}=0.5$.}\label{figg3}
\end{figure}

In spite of the weak dependence with the qubit-qubit separation, there exist nevertheless specific separations which are optimum for the entanglement generation. As \autoref{figg2}c shows, the concurrence $C_{\text{max}}$ is maximized when the distance between the qubits is $2\tilde{d} = 0,1,2...$ For these separations, an analytical expression can be extracted from \autoref{conc},
\begin{equation}\label{Cmaxt}
C_{\text{max}}^2 = \frac{(1+\Delta_1)(1+\Delta_2)}{1 - q^2} \left(\frac{1 - q}{1+q}\right)^{\frac{1}{q}}.
\end{equation}
To understand the dependence of $C_{\text{max}}$ with the directionalities, which is shown in \autoref{figg3}, it is important to bear in mind that initially only qubit $1$ is excited. For $\Delta_j = -1$ the concurrence is strictly zero during all the time evolution, as the qubit $j$ is coupled only to left-propagating modes and thus is not able to interact with its partner. On the other hand, any values of the directionalities in the region $\Delta_1,\Delta_2 >0$ result in an enhancement of $C_{\text{max}}$ with respect to the non-chiral case. Moreover, when both qubits are maximally coupled to right-propagating modes, i.e., $\Delta_1 = \Delta_2 = 1$, the maximum entanglement rises up to a very large value, which can be extracted from \autoref{Cmaxt} as $\lim_{q\to 0} C_{\text{max}} = 2/e \sim 0.73$. This is a significant result, as it shows that the maximum achievable concurrence can be enhanced by a factor of $\sim 50\%$ with respect to the non-chiral coupling scheme.

We have seen that playing upon chirality it is possible both to reach higher values of the maximum concurrence, $C_{\text{max}}$, and to reduce the sensitivity to the qubits separation, $d$. However, for large separations or coupling rates, $\gamma d\gg v_g$, it has been reported that non-Markovian effects arise in waveguide systems \cite{NJP}. Such effects introduce additional retardation that cannot be described with the currently used master equation formalism. Hence, we now reformulate the problem to properly assess the robustness of the proposed protocol. We employ a more complete approach by diagonalizing the full Hamiltonian of the system in the single excitation subspace, as detailed in the supplemental material \cite{suppl}.  Unless stated otherwise, we will assume equally coupled qubits, i.e., $\gamma_{1R} = \gamma_{2R}$ and $\gamma_{1L} = \gamma_{2L}$ and large directionalities ($\Delta_1 = \Delta_2 = 0.90$) in order to stay close to the optimum configuration. We also include explicitly the lossy modes in this part of our work, by fixing the decay rates $\Gamma_j$ such that $\beta_j = 0.98$. Note that the chosen values for both the directionalities and the beta factors have been experimentally reported \cite{LodahlarxiV3}.

\begin{figure}
	\centering
	\includegraphics[width=\linewidth]{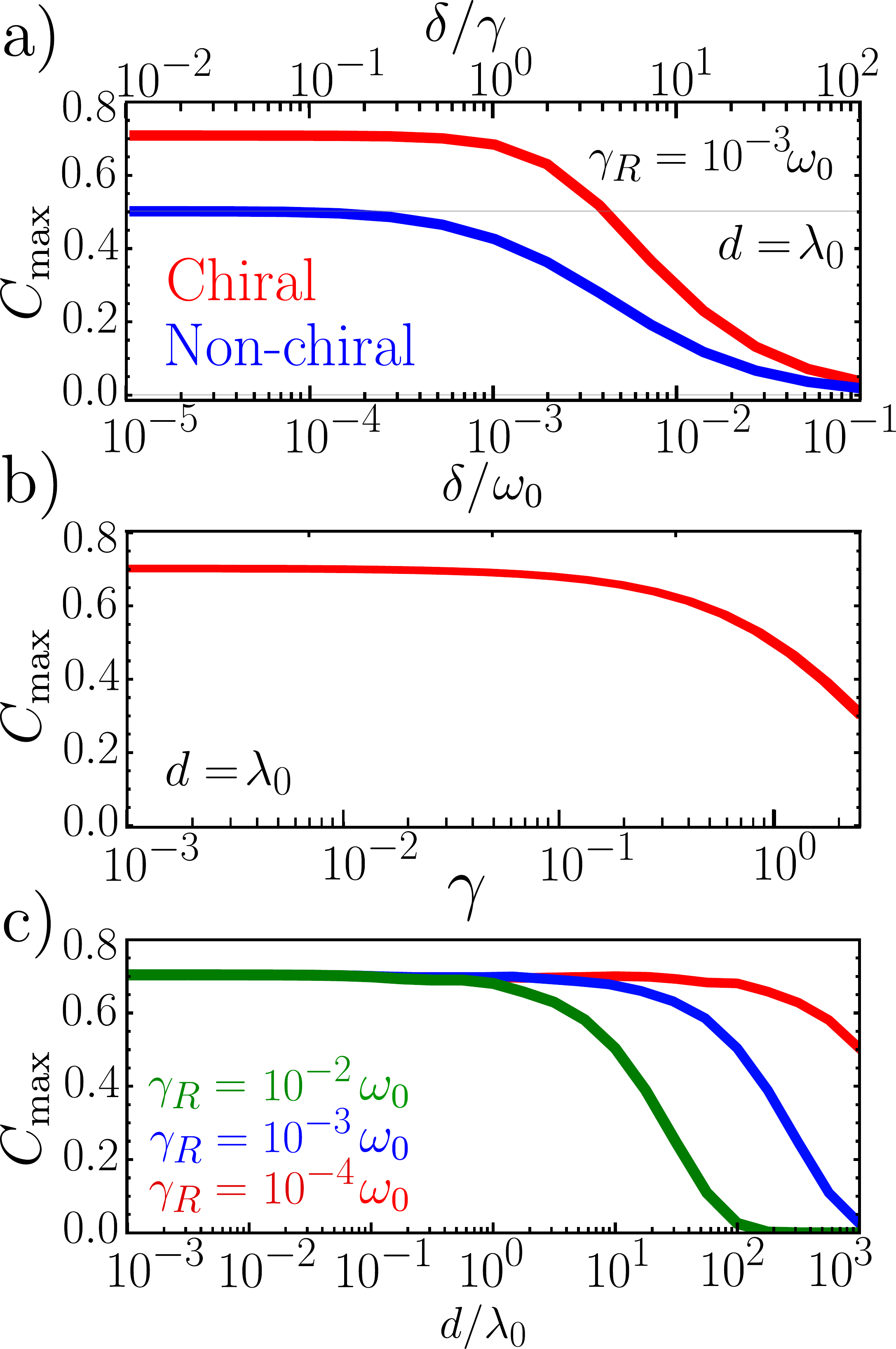}
	\caption{Maximum concurrence $C_{\text{max}}$ as a function of the various system parameters $(\Delta_{j} = 0.9,\beta_j = 0.98)$. a) Effect of the detuning between the transition frequencies of the qubits, $\delta$. b) Dependence on the total coupling to the waveguide, $\gamma$. c) Dependence on the qubit-qubit separation $d$, for different values of $\gamma$. All the coupling constants are normalized to $\omega_0$.}\label{figg4}
	\vspace{0cm}
\end{figure}

We employ the above mentioned formalism to explore the robustness of the scheme against variation of three parameters, namely the detuning between the frequencies of the emitters, the total coupling to the waveguide modes, $\gamma$, and the separation $d$. In \autoref{figg4}a, the effect of detuning between the qubits is studied. For the corresponding calculations, the frequencies of both emitters are modified according to $\omega_1 =\omega_0 + \delta/2$, and $\omega_2 =\omega_0 - \delta/2$. Physically, we are shifting away the emission spectrum of qubit $1$ from the absorption spectrum of qubit $2$, keeping their linewidth constant. It is known that a strong overlap between both the spectral distribution of the photon emitted by qubit $1$, and the absorption spectrum of qubit $2$, is key for entanglement generation \cite{GonzalezBallPRA2014}. Hence, the concurrence naturally decreases for large values of $\delta$. The detuning relative to $\omega_0$ seems to be the most critical parameter as a change of $\sim 0.5 \%$ in the frequency of the qubits is enough to reduce $C_{\text{max}}$ below $0.5$. However, the robustness against detuning is considerably large with respect to the qubits linewidth, $\gamma$. Indeed, whereas in the non-chiral case the
concurrence is independent on the detuning  for $\delta \lesssim 0.2 \gamma$, for chiral couplings this range is increased by a factor of $\sim5$. Additionally, in the chiral case, concurrences of $C_{\text{max}} = 0.5$ are possible for detunings as large as $\delta \sim 5\gamma$. This is a crucial advantage with respect to non-chiral systems, especially for quantum emitters with a very narrow linewidth such as quantum dots.

The variation of $C_{\text{max}} $ with the total qubit-waveguide coupling is displayed in \autoref{figg4}b. For low values of $\gamma$, the concurrence is close to its theoretical maximum $\left(C_{\text{max}} \sim 0.7\right)$ due to the large chosen directionalities. When $\gamma$ is increased, the photonic wavepacket emitted by qubit $1$ becomes narrower in space\cite{GonzalezBallPRA2014} and, eventually, its width becomes smaller or comparable to the qubit-qubit separation $d$. As a result, the qubit $1$ significantly decays before the photon reaches qubit $2$ and, at any given time, at least one of the emitters is fairly depopulated. Thus, the concurrence $C = 2\vert\rho_{12}\vert = 2\sqrt{\rho_{11}\rho_{22} }$ decreases for large values of $\gamma$. This effect has been studied in detail in non-chiral configurations \cite{NJP}. Note that, nevertheless, even for couplings as large as $\gamma \approx 0.1\omega_0$, the concurrence remains above $0.6$. This result shows that chirality allows for a high level of concurrence not only in the optical regime, but also in systems where much larger couplings arise such as superconducting stripes.
 
 Finally, the variation of the concurrence with the qubit-qubit separation $d$ is shown in \autoref{figg4}c. While in the non-chiral case a maximum concurrence of $0.5$ was obtained only for particular values of $d$, for chiral couplings the entanglement generation scheme is shown to be robust for a wide range of separations. For large distances there is a decay in the concurrence, which responds to the same mechanism discussed above. In this case, the spatial extension of the emitted photon is made smaller than the separation $d$ by directly increasing the qubit separation instead of the coupling $\gamma$. Interestingly, for qubit-waveguide couplings in the optical regime $\left(\gamma \lesssim 10^{-4}\omega_0\right)$, the entanglement generation scheme is extremely robust with respect to the distance $d$, allowing for concurrences above $0.6$ for very large separations, e.g. around  $60~\mathrm{\mu m}$ at $\omega_0 \sim 2~\mathrm{eV}$. The separation between the qubits is thus not a critical parameter anymore, allowing for a much easier implementation of this entanglement protocol.
 
 In conclusion, the phenomenon of spontaneous generation of entanglement between two qubits chirally coupled to a waveguide has been analyzed in detail. We show that even the slightest directionalities in the couplings may improve the maximum achievable entanglement, as compared to non-chiral systems. Moreover, we identify the optimal directionalities and demonstrate a very significant enhancement of the maximum entanglement. This entangling scheme displays a fairly weak dependence on the relevant parameters, which highlights the robustness of the protocol. In particular, when compared to the non-chiral case, the influence of the qubit-qubit separation is reduced, which constitutes an important advantage for the feasibility of an experimental implementation. In addition to the interest of our specific results, our work positions entanglement as one further reason for the exploration of chiral waveguide QED.

\begin{acknowledgments}
CGB acknowledges the Spanish MECD (FPU13/01225 fellowship). AGT acknowledges support from the Intra-European Fellowship NanoQuIS (625955) and the European Union integrated project Simulators and Interfaces with Quantum Systems (SIQS). CGB and FJGV acknowledge the European Research Council (ERC-2011-AdG Proposal No. 290981). EM, FJGV and CGB acknowledge the Spanish MINECO (MAT2014-53432-C5-5-R grant).
\end{acknowledgments}

\bibliography{bibliography}

\pagebreak

\onecolumngrid
\begin{center}
	\textbf{\large A chiral route to spontaneous entanglement generation: Supplemental Material}
\end{center}
\setcounter{equation}{0}
\setcounter{figure}{0}
\setcounter{table}{0}
\setcounter{page}{1}
\makeatletter

\vspace{0.9cm}

\twocolumngrid

\section*{Diagonalization of the Hamiltonian.}
We make use of the real-space formalism which is usually employed in low-excitation problems in waveguide QED \cite{FanPRA2009}. The Hamiltonian of the system is $H = H_{qb} + H_{wg} + H_{I}$, where the two first terms correspond to the energy of the qubits and the waveguide, respectively, and $H_I$ is the interaction term. They are given by $\left( \hbar = 1\right)$:
\begin{align}\label{Hamilt}
	& H_{qb} = \sum_j \Omega_j\sigma_j^\dagger\sigma_j,\\
	& H_{wg} = iv_g\!\int\!\! dx \left[ c^{\dagger}_L (x)\partial_xc_L(x)-c^{\dagger}_R(x)\partial_xc_R(x)\right]\!,\\
	& H_{I} \!=\! \sum_{j=1}^2\!\sum_{\alpha=R,L}\!\int\!\! dx \delta(x-x_j)\!\left[V_{j\alpha}c_\alpha^\dagger(x)\sigma_j + h.c.\right]\!.\!
\end{align}
In the above equations, $\Omega_j$ is the transition frequency of qubit $j$, and $v_g$ is the group velocity of the guided modes, whose dispersion is considered linear. The constants $V_{j\alpha}$, assumed real for simplicity, are related to the coupling rates in the main text through $\gamma_{j\alpha} = V_{j\alpha}^2/v_g$. The operators $\sigma_j$ and $c_\alpha(x)$ destroy an excitation in qubit $j$ and a $\alpha-$ propagating photon at position $x$, respectively. Finally, $x_j = \pm d/2$ is the position of the emitter $j$ along the waveguide. Note that the losses $\Gamma_j$ are not accounted for in \autoref{Hamilt}, as we introduce them a posteriori.

In order to diagonalize the system Hamiltonian in the single excitation subspace, we assume an eigenstate of the form
\begin{equation}
	\vert \epsilon \rangle = \left(\sum_j \alpha_j \sigma_j^\dagger + \sum_{\alpha}\int dx \phi_\alpha(x) c^\dagger_\alpha(x)\right)\vert 0 \rangle,
\end{equation}
and solve the time-independent Schr\"odinger equation, $H \vert \epsilon \rangle = \epsilon \vert \epsilon \rangle$, to obtain the coefficients $\lbrace \alpha_j, \phi_\alpha(x)\rbrace$. Following the usual approach, we make a plane wave Ansatz for the wavefunctions $\phi_\alpha(x)$, i.e.
\begin{equation}\label{Ans1}
	\phi_R(x) = e^{i\epsilon x/v_g} \Bigg\lbrace\begin{array}{lll} A & \;\; \text{for} \; & x<-d/2\\
		B &  \;\; \text{for} \; & -d/2<x<d/2\\
		C & \;\; \text{for} \; & d/2<x
	\end{array},
\end{equation}
\vspace{-0.2cm}
\begin{equation}\label{Ans2}
	\phi_L(x) = e^{-i\epsilon x/v_g} \Bigg\lbrace\begin{array}{lll} D & \;\; \text{for} \; & x<-d/2\\
		E &  \;\; \text{for} \; & -d/2<x<d/2\\
		F & \;\; \text{for} \; & d/2<x
	\end{array},
\end{equation}
which reduces the problem to an algebraic system of equations \cite{FanPRA2009}. For each energy $\epsilon$, we can naturally find two linearly independent eigenstates, corresponding to the scattering of photons coming from either $x = -\infty$ or $x = \infty$. These states, labeled $\vert \epsilon_+\rangle$ and $\vert \epsilon_-\rangle$, are obtained by setting $\lbrace A\!=\!1,F\!=\!0\rbrace$ and $\lbrace F\!=\!1,A\!=\!0\rbrace$ in Eqs \ref{Ans1}-\ref{Ans2}, respectively. The eigenstates $\vert \epsilon_\pm\rangle$, whose degeneracy is inherited from that of the right/left propagating modes of a bare waveguide, have been called \textit{scattering eigenstates} in the literature.

The scattering solutions of Schr\"odinger equation have been reported to not always form a complete basis, as localized resonances may arise \cite{NJPcopy}. We thus need to check our equations for orthogonal solutions, i.e. $A=F=0$. A simple calculation shows that the necessary conditions for a localized eigenstate are $\gamma_{1R} = \gamma_{1L}  $ and $ \gamma_{2R} = \gamma_{2L}$. We conclude that, in the chiral case, the basis $\lbrace \vert \epsilon_\pm \rangle\rbrace $ is complete, as no localized resonances appear.

With the eigenstates at hand, we are able to introduce the qubit losses, $\Gamma_j$. As localized eigenstates are not present in the chiral case, the effect of the lossy modes is easily accounted for. Indeed, it has been shown in ref. \cite{NJPcopy} that the effect of the external decay of the qubits over the scattering eigenstates can be fully reproduced by adding an imaginary part to the frequency of the emitters, $\Omega_j \to \Omega_j - i \Gamma_j/2$.

The final step for obtaining the system dynamics is to construct the time evolution operator, $U(t)$. If the coupling is chiral, this task is not as straightforward as in previous works, as the two scattering branches are not orthogonal. The operator $U(t)$ takes now a more complicated form:
\begin{equation}\label{Ut}
	U(t) = \frac{1}{2\pi v_g}\sum_{i,j=\pm}\int_{-\infty}^{\infty} d\epsilon \; e^{-i\epsilon t} \vert \epsilon_i \rangle \left(S^{-1}\right)_{ij} \langle \epsilon_j \vert,
\end{equation}
where we define the overlap matrix $S$ as 
\begin{equation}
	S_{ij} = \lim_{L\to\infty}\frac{\langle\epsilon_i \vert \epsilon_j \rangle}{L}.
\end{equation}
Using \autoref{Ut} we can numerically obtain the evolution of any initial state through $\vert \psi(t) \rangle = U(t) \vert \psi(0)\rangle$. Finally, it can be checked that $U(0) = 1$, which certifies the completeness of the basis $\lbrace \vert \epsilon_\pm \rangle\rbrace $.

\end{document}